\documentclass[conference]{IEEEtran}

\usepackage{graphicx}

\ifCLASSINFOpdf
\else
\fi

\begin{document}

\title{Cooperative Interference Control for Spectrum Sharing in OFDMA Cellular Systems}

\author
{ \IEEEauthorblockN{Bin Da${}^\dagger$} \IEEEauthorblockA
{${}^\dagger$ECE Department, National University of Singapore\\
Email:\{dabin,elezhang\}@nus.edu.sg} \and \IEEEauthorblockN{Rui
Zhang${}^\dagger{}^\ddagger$} \IEEEauthorblockA
{${}^\ddagger$Institute for Infocomm Research, A*STAR, Singapore\\
Email: rzhang@i2r.a-star.edu.sg}}

\maketitle

\begin{abstract}
This paper studies cooperative schemes for the inter-cell
interference control in
orthogonal-frequency-division-multiple-access (OFDMA) cellular
systems. 
The downlink transmission in a simplified two-cell system is
examined, where 
both cells simultaneously access
the same frequency band using OFDMA. 
The joint power and subcarrier allocation over the two cells is
investigated for maximizing their sum throughput with both
centralized and decentralized implementations. Particularly, the
decentralized allocation is achieved via a new \emph{cooperative
interference control} approach, whereby the two cells independently
implement resource allocation to maximize individual throughput in
an iterative manner, subject to a set of 
mutual interference power constraints. 
Simulation results show that the proposed
decentralized resource allocation schemes achieve the system
throughput close to that by the centralized scheme, and provide
substantial throughput gains over existing schemes.
\end{abstract}



%
\IEEEpeerreviewmaketitle

\section{Introduction}\label{section1}

In traditional cellular networks, the base stations (BSs) in
different cells independently control the transmission with their
associated users. The inter-cell interference is avoided or
minimized by adopting different frequency reuse patterns, which only
allow non-adjacent cells to reuse the same frequency band. The
frequency reuse factor is assigned to specify the rate at which the
same frequency band can be used in the network. Due to emerging
high-rate wireless multimedia applications, traditional cellular
systems have been pushed towards their throughput limits. As a
result, it has been proposed to increase the frequency reuse factor
such that each cell can be assigned with more frequency bands to
increase the attainable throughput. In the special case where all
cells can share the same frequency band for simultaneous
transmission, this corresponds to the factor-one or {\it universal}
frequency reuse. However, with more flexible frequency reuse, the
inter-cell interference control becomes an essential problem in
cellular systems, which has recently drawn significant research
attentions (see, e.g., \cite{Gesb10}$-$\cite{Zhang10new}). 

For multicell systems with a universal frequency reuse, two
promising approaches have been proposed to resolve the inter-cell
interference problem (see, e.g., \cite{Gesb10} and the references
therein): {\it interference coordination} and {\it network MIMO
(multiple-input multiple-output)}. In the former approach, the
performance of a multicell system is optimized via joint resource
allocation among all cells, based on their shared channel state
information (CSI) of all direct and interfering links across
different cells. Furthermore, if the baseband signal synchronization
among the BSs of different cells is available and the transmit
messages of different cells are shared by their BSs, a more powerful
cooperation can be achieved in the downlink via jointly encoding the
transmit messages of all BSs. In this so-called network MIMO
approach, the combined use of antennas at different BSs for joint
signal transmission resembles the conventional single-cell
multiantenna broadcast channel (BC) \cite{Gesb10}. In this paper,
the former interference coordination approach is adopted due to its
relatively easier implementation in practical systems.

More specifically, we study the inter-cell
interference coordination for a two-cell OFDMA downlink system with
universal frequency reuse. All BSs and user terminals are assumed to
be each equipped with a single antenna, and thus the system of
interest can be modeled as a {\it parallel interfering SISO
(single-input single-output) BC}. Promising applications of this
two-cell system model are illustrated in Fig. \ref{fig.1}, which
shows a geographically symmetric setup with two adjacent macrocells,
as well as a non-symmetric setup with one macrocell and one inside
femtocell. This paper investigates the joint power and subcarrier
allocation over the two cells to maximize their sum throughput, for
both centralized and decentralized implementations. Specifically,
for the centralized allocation, with the assumption of a global
knowledge of all channels in the network, we propose a scheme to
jointly optimize power and subcarrier allocation over the two cells
by applying the {\it Lagrange duality} method from convex
optimization \cite{Boy04}. This centralized scheme provides a
performance benchmark for the decentralized schemes studied
subsequently.

For the decentralized resource allocation, this paper proposes a new
{\it cooperative interference control} approach, whereby the two
cells independently optimize resource allocation to maximize
individual throughput subject to a set of preassigned mutual
interference power constraints, in an iterative manner until the
resource allocation in both cells converges. Two types of
interference power constraints are further examined: one is to
constrain the total interference power across all subcarriers from
each cell to the active users in its adjacent cell, termed
\emph{joint subcarrier protection} (JSP); and the other is to limit
the interference power over each individual subcarrier, termed
\emph{individual subcarrier protection} (ISP). Also, the optimal
resource allocation rules for each cell to maximize individual
throughput with JSP or ISP are derived.


The rest of this paper is organized as follows. Section II
introduces the two-cell downlink OFDMA system, and formulates the
optimization problem for resource allocation. Section III presents
the centralized resource allocation scheme. Section IV proposes two
decentralized schemes via the cooperative interference control
approach with JSP and ISP, respectively. Section V presents
simulation results and pertinent discussions. Finally, Section VI
concludes the paper.

\section{System Model And Problem Formulation}\label{section2}

\begin{figure}[!t]
  \centering
  \includegraphics[width=75mm,height=50mm]{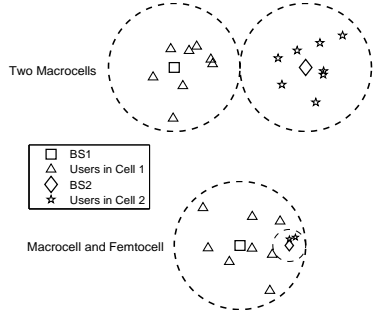} \\
  \vspace{-16pt}
  \caption{System model for two-cell applications.}
  \label{fig.1}\vspace{-8pt}
\end{figure}


As shown in Fig. \ref{fig.1}, we consider a two-cell system sharing
the same frequency band with each cell having a downlink OFDMA
transmission. We use $m \in \left\{ {1,2} \right\}$ to denote each
of the two cells, which are referred to as the 1st and 2nd cells in
this paper, respectively. For convenience, let the 1st cell refer to
the macrocell and the 2nd cell refer to either the macrocell or the
femtocell in Fig. \ref{fig.1}. The total system bandwidth shared by
the two cells is assumed to be $B$ Hz, which is equally divided into
$N$ subcarriers (SCs) indexed by $n \in \Lambda {=} \left\{
{1,2,...,N} \right\}$. Each SC is assumed to be used by at most one
user inside each cell and could be shared between two users
individually selected from the two cells. In addition, the users in
the network are indexed by ${k_1} \in {\Delta _1}{=}\left\{
{1,2,...,{K_1}} \right\}$ in the 1st cell and ${k_2} \in {\Delta
_2}{=}\left\{ {1,2,...,{K_2}} \right\}$ in the 2nd cell, where $K_1$
and $K_2$ are the total numbers of users in each corresponding cell.

Furthermore, we denote the channel power gains (amplitude squares)
from the two BSs to their respective users, saying users $k_1, k_2$,
in each cell as ${h_{n{k_1}}}$ and ${h_{n{k_2}}}$, respectively. The
inter-cell interference channel gain from $\rm{BS}_1$ to $k_2$ is
denoted by ${g_{n{k_2}}}$, while that from $\rm{BS}_2$ to $k_1$ is
by ${g_{n{k_1}}}$. We assume that the noise at each user's receiver
has independent circularly symmetric complex Gaussian (CSCG)
distribution over SCs with zero mean and variance ${\sigma ^2} =
{{{z_0}B} \mathord{\left/ {\vphantom {{{z_0}B} N}} \right.
\kern-\nulldelimiterspace} N}$, denoted by $\mathcal{CN}\left(
0,{\sigma ^2} \right)$, where $z_0$ is the noise power spectral
density. In addition, 
the transmit power allocated to user $k_1$ at SC $n$ is denoted by
${p_{n{k_1}}}$. Thus, over all users and SCs in the 1st cell, we can
define a power allocation matrix ${{\bf{P}}_1}$ ($K_1$-by-$N$) with
the non-negative elements denoted by ${p_{n{k_1}}}, n \in \Lambda
,{k_1} \in {\Delta _1}$. ${{\bf{P}}_1}$ is assumed to satisfy an
OFDMA-based power allocation (OPA), in which there exists at most
one element in each column being larger than zero and all the other
elements are equal to zero. This OPA constraint can be expressed as
${{\bf{P}}_1} \in {S_1} = \left\{ {{{\bf{P}}_1} \succeq 0 | ~
{p_{n{k_1}}}{p_{n{{k'}_1}}} = 0,\forall {k_1} \ne {{k'}_1},\forall
n} \right\}$. Similarly, we can define the power allocation matrix
for the 2nd cell as ${{\bf{P}}_2}\in {S_2}$ ($K_2$-by-$N$ matrix)
under a similar OPA constraint. 
The $n$th columns of
${{\bf{P}}_1}$ and ${{\bf{P}}_2}$ are denoted by two vectors
${{\bf{p}}_{1n}} \in {{S}_{1n}}$ and ${{\bf{p}}_{2n}} \in
{{S}_{2n}}$, where ${{S}_{1n}}$ (${{S}_{2n}}$) is drawn from the
$n$th column of ${S_1}$ (${S_2}$).

With the above system model and assuming that the inter-cell
interference is treated as additional Gaussian noise at each user's
receiver, the signal-to-interference-plus-noise-ratio (SINR) of user
$k_1$ at SC $n$ in the 1st cell is given by
\begin{equation}\label{equ.1}
SIN{R_{n{k_1}}} =
\frac{{{p_{n{k_1}}}{h_{n{k_1}}}}}{{\sum\nolimits_{{k_2} = 1}^{{K_2}}
{{p_{n{k_2}}}{g_{n{k_1}}}}  + {\sigma ^2}}}.
\end{equation}
Similarly, the SINR of user $k_2$ at SC $n$ in the 2nd cell is
denoted by $SIN{R_{n{k_2}}}$.
Thus, the achievable sum-rate of user ${k_m} \in {\Delta _m}, m
\in \left\{ {1,2} \right\}$ is given by 
\begin{equation}\label{equ.4}
{r_{{k_m}}} = \frac{1}{N}\sum\nolimits_{n = 1}^N {{\log _2}\left( {1
+ SIN{R_{n{k_m}}}} \right)}.
\end{equation}
We consider the weighted-sum-rate (WSR) in each cell i.e., 
\begin{equation}\label{equ.5}
{R_m} = \sum\nolimits_{{k_m} = 1}^{{K_m}} {{w_{{k_m}}}{r_{{k_m}}}}
,{\rm{ }} m \in \{ 1,2\},
\end{equation} where ${w_{{k_m}}}$ is the (non-negative) rate weight of user $k_m$ in the $m$th cell.
With individual transmission power constraint at each BS, 
the following optimization problem can be formulated to maximize the
{\it system throughput} defined as
\begin{equation}\label{equ.6}
\mathop {\max }\limits_{{{\bf{P}}_1},{{\bf{P}}_2}} {R_1} + {R_2}
\end{equation}
\begin{equation}\label{equ.7}
{\rm subject~to~}\sum\nolimits_{{k_m} = 1}^{{K_m}} {\sum\nolimits_{n
= 1}^N {{p_{n{k_m}}}}  \le P_m^{{\rm{BS}}}} ,{\rm{ }} m \in \{
1,2\},
\end{equation} where $P_m^{{\rm{BS}}}$ is the given power constraint at ${\rm BS}_m$,
and
\begin{equation}\label{equ.8}
{{\bf{P}}_m} \in {S_m},{\rm{ }} m \in \{ 1,2\}
\end{equation} is the OPA constraint for the $m$th cell.


\section{Centralized Allocation}\label{section3}

In this section, we study the centralized optimization for jointly
allocating resources in the two cells so as to maximize the system
throughput, which corresponds to solving Problem (\ref{equ.6})
globally with constraints (\ref{equ.7}) and (\ref{equ.8}). For the
centralized allocation, it is assumed that all channel gains in the
network are collected by a central controller, which is capable of
performing a centralized resource allocation and informing the
allocation results to each cell for data transmission.

Due to the non-convex OPA constraint and the non-concave objective
function over ${{\bf{P}}_1}$ and ${{\bf{P}}_2}$, the optimization
problem in (\ref{equ.6}) is {\it non-convex} and thus cannot be
solved efficiently for the global optimum. Nevertheless, the
Lagrange duality method \cite{Boy04} can be applied to this problem
to obtain a suboptimal solution. Interestingly, according to
\cite{Yu05}, it has been shown that a so-called ``time-sharing''
condition usually holds for resource allocation problems in OFDMA,
and the duality gap for such problems solved by the Lagrange duality
method becomes asymptotically zero as the number of subcarriers in
the system becomes large. Accordingly, in the sequel, we apply the
Lagrange duality method to solve Problem
(\ref{equ.6}).\footnote{Based on numerical results, the duality gap
for our problem at hand is nonzero only for a negligibly small
portion of the total number of randomly generated channels, even
with a not-so-large number of subcarriers.}

First, we express the {\it partial} Lagrangian of Problem
(\ref{equ.6}) as
\begin{equation}\label{equ.9}
\begin{array}{l}
L\left( {{{\bf{P}}_1},{{\bf{P}}_2},{\lambda _1},{\lambda _2}}
\right) = \sum\nolimits_{n = 1}^N {{L_n}\left(
{{{\bf{p}}_{1n}},{{\bf{p}}_{2n}},{\lambda _1},{\lambda _2}} \right)}  \\
{\rm{\ \ \ \ \ \ \ \ \ \ \ \ \ \ \ \ \ \ \ \ \ \ \ \ }}
+ {\lambda _1}P_1^{{\rm{BS}}} + {\lambda _2}P_2^{{\rm{BS}}}, \\
\end{array}
\end{equation} where, for each SC $n \in \Lambda$,
\setlength{\arraycolsep}{0.0em}
\begin{equation}\label{equ.10}
\begin{array}{l}
 {L_n}\left( {{{\bf{p}}_{1n}},{{\bf{p}}_{2n}},{\lambda _1},{\lambda _2}} \right) = \sum\nolimits_{{k_1} = 1}^{{K_1}} {{w_{{k_1}}}{r_{n{k_1}}}}  \\
+ \sum\nolimits_{{k_2} = 1}^{{K_2}} {{w_{{k_2}}}{r_{n{k_2}}}}  - {\lambda _1}\sum\nolimits_{{k_1} = 1}^{{K_1}} {{p_{n{k_1}}}}  - {\lambda _2}\sum\nolimits_{{k_2} = 1}^{{K_2}} {{p_{n{k_2}}}}, \\
 \end{array}
\end{equation}
and ${\lambda _1},{\lambda _2}$ are non-negative dual variables
associated with the power constraints in (\ref{equ.7}) with $m=1$
and $2$, respectively. The Lagrange dual function is then given by
\begin{equation}\label{equ.11}
g\left( {{\lambda _1},{\lambda _2}} \right) = \mathop {\max
}\limits_{{{\bf{P}}_1} \in {S_1},{{\bf{P}}_2} \in {S_2}} L\left(
{{{\bf{P}}_1},{{\bf{P}}_2},{\lambda _1},{\lambda _2}} \right).
\end{equation}
Hence, the dual problem can be defined as
\begin{equation}\label{equ.12}
\mathop {\min }\limits_{{\lambda _1} \ge 0,{\lambda _2} \ge 0}
g\left( {{\lambda _1},{\lambda _2}} \right).
\end{equation}
For a given pairs of ${\lambda _1}$ and ${\lambda _2}$, we have
\begin{equation}\label{equ.13}
g\left( {{\lambda _1},{\lambda _2}} \right) = \sum\nolimits_{n =
1}^N {g_n^{}\left( {{\lambda _1},{\lambda _2}} \right)}  + {\lambda
_1}P_1^{{\rm{BS}}} + {\lambda _2}P_2^{{\rm{BS}}},
\end{equation}
where $g_n^{}\left( {{\lambda _1},{\lambda _2}} \right), n \in
\Lambda$, is obtained by solving the following per-SC maximization
problem
\begin{equation}\label{equ.14}
g_n^{}\left( {{\lambda _1},{\lambda _2}} \right) = \mathop {\max
}\limits_{{{\bf{p}}_{1n} \in {{S}_{1n}} },{{\bf{p}}_{2n} \in
{{S}_{2n}} }} {L_n}\left( {{{\bf{p}}_{1n}},{{\bf{p}}_{2n}},{\lambda
_1},{\lambda _2}} \right).
\end{equation}

The maximization problem in (\ref{equ.11}) is thus decoupled into
$N$ per-SC resource allocation problems given by (\ref{equ.14}). 
Due to the OPA constraints, for one particular SC $n$, it can be
simultaneously assigned to one pair of users $\left( {{k_1},{k_2}}
\right)$ from the two cells when the resultant ${L_n}\left(
{{{\bf{p}}_{1n}},{{\bf{p}}_{2n}},{\lambda _1},{\lambda _2}} \right)$
in (\ref{equ.10}) attains its maximum value (with the optimized
$p_{nk_1}$ and $p_{nk_2}$). This user pair can be obtained by
searching over all possible combinations from users ${k_1} \in
{\Delta _1}, {k_2} \in {\Delta _2}$. Thus, the optimal SC and power
allocation that solves the problem in (\ref{equ.14}) is
\begin{equation}\label{equ.15}
\left( {{{\bar k}_1},{{\bar k}_2}} \right) = \mathop {\arg \max
}\limits_{{k_1} \in {\Delta _1},{k_2} \in {\Delta _2}} \left\{
{\mathop {\max }\limits_{{p_{n{k_1}}} \ge 0,{p_{n{k_2}}} \ge 0}
L_n^{\left( {{k_1},{k_2}} \right)}} \right\},
\end{equation} where $\left( {{{\bar k}_1},{{\bar k}_2}} \right)$ is
the selected user pair to share SC $n$, and
\begin{equation}\label{equ.16}
L_n^{\left( {{k_1},{k_2}} \right)} = {w_{{k_1}}}{r_{n{k_1}}} +
{w_{{k_2}}}{r_{n{k_2}}} - {\lambda _1}{p_{n{k_1}}} - {\lambda
_2}{p_{n{k_2}}}
\end{equation} is obtained from (\ref{equ.10}).
For a given pair of $\left( {{k_1},{k_2}} \right)$, the optimal
${p_{n{k_1}}}$ and ${p_{n{k_2}}}$ to maximize $L_n^{\left(
{{k_1},{k_2}} \right)}$ in (\ref{equ.16}) have no closed-form
solutions due to the non-convexity of this problem. However, an
iterative search based on, e.g., Newton's method \cite{Boy04} can be
utilized to find a pair of local optimal solutions for
${p_{n{k_1}}}$ and ${p_{n{k_2}}}$. 
Then, we can check all possible user combinations to determine the
optimal SC allocation according to (\ref{equ.15}) with optimized
power allocation. 

After solving the per-SC problems in (\ref{equ.14}) for all $n$'s, a
subgradient-based method, e.g., the ellipsoid method, can be adopted
to solve the dual problem in (\ref{equ.12}) so that the power
constraints in (\ref{equ.7}) at both BSs are satisfied.
The details are thus omitted for brevity. 
Note that Problem (\ref{equ.6}) can be solved in polynomial time
with an overall complexity with order $O\left(
{Itr{_{out}}N{K_1}{K_2}Itr{_{in}}} \right)$ in its dual domain.
Specifically, for one particular SC, we search for $K_1K_2$
combinations of user pairs and determine the power allocation for
each user pair with $Itr{_{in}}$ iterations. In addition,
$Itr{_{out}}$ is the number of iterations for solving the dual
problem in (\ref{equ.12}). However, this centralized allocation
needs a system level coordination with
all channel conditions in the two cells, which is a
demanding requirement for practical applications. In the next
section, we propose decentralized schemes for resource allocation,
which can be implemented by each cell independently.

\section{Decentralized Allocation}\label{section4}

In this section, a new cooperative interference control approach is
applied to design decentralized resource allocation schemes for the
two-cell OFDMA downlink system. In this approach, each cell
independently optimizes its resource allocation to maximize
individual WSR under its BS's own transmit power constraint, as well
as a set of newly imposed constraints to regulate the leakage
interference power levels to the active users in its adjacent cell.
The above operation iterates between the two cells, until both cells
obtain a converged resource allocation under their mutual inter-cell
interferences. Specifically, two decentralized allocation schemes
are studied in this section corresponding to two different types of
interference power constraints, namely JSP and ISP.

\subsection{Joint Subcarrier Protection (JSP)}

In this subsection, we solve the optimal resource allocation problem
of maximizing the 1st cell's WSR subject to its BS's power
constraint and a given JSP constraint to the active users in the 2nd
cell. Similar problem formulation and solution apply to the resource
allocation in the 2nd cell and are thus omitted. 

Consider the resource allocation problem in the 1st cell subject to
the leakage interference constraint for the 2nd cell. In order to
characterize the leakage interference to the 2nd cell, $\rm{BS}_1$
needs to know the interference channel gains from it to all active
users over different SCs in the 2nd cell. Let ${\bar k_2} = \pi _2^n
\in {\Delta _2}$ denote the active user at SC $n$ in the 2nd cell,
with the corresponding interference channel gain from $\rm{BS}_1$ to
${\bar k_2}$ being ${g_{n{{\bar k}_2}}}$. It is then assumed that
${g_{n{{\bar k}_2}}}$ has been perfectly estimated by user ${\bar
k_2}$ in the 2nd cell and fed back to $\rm{BS}_2$. After collecting
${g_{n{{\bar k}_2}}}$ for all $n$'s from its active users,
$\rm{BS}_2$ sends these channel gain values to $\rm{BS}_1$ (via a
backhaul link connecting these two BSs).
Note that if a particular
SC $n$ is not used by any user in the 2nd cell, the corresponding
interference channel gain ${g_{n{{\bar k}_2}}}$ sent from
$\rm{BS}_2$ to $\rm{BS}_1$ is set to be zero regardless of its
actual value, so that this SC can be used by the 1st cell without
any interference constraint.

To maximize the WSR of the 1st cell, the following problem is
formulated as
\begin{equation}\label{equ.20}
\mathop {\max }\limits_{{{\bf{P}}_1} \in {S_1}} {R_1}
\end{equation} subject to
\begin{equation}\label{equ.21}
\sum\nolimits_{{k_1} = 1}^{{K_1}} {\sum\nolimits_{n = 1}^N
{{p_{n{k_1}}}}  \le P_1^{{\rm{BS}}}},
\end{equation}
\begin{equation}\label{equ.22}
\frac{1}{N}\sum\nolimits_{n = 1}^N {\sum\nolimits_{{k_1} =
1}^{{K_1}} {{p_{n{k_1}}}{g_{n{{\bar k}_2}}}} }  \le {T_2},
\end{equation}
where $T_2$ is the given JSP power constraint for protecting all the
active users in the 2nd cell. Note that $T_2$ limits the
interference power averaged over all the SCs; thus, the
corresponding resource allocation scheme is refereed to as the {\bf
Average} scheme for convenience.


We assume the non-negative dual variables associated with
(\ref{equ.21}) and (\ref{equ.22}) are $\lambda ,\mu$. Similarly as
in the case of centralized allocation, for a given pair of $\lambda,
\mu$, Problem
(\ref{equ.20}) can be decoupled into $N$ per-SC problems in its dual domain, 
and the optimal 
allocation for SC $n$ in the 1st cell is derived as 
\begin{equation}\label{equ.27}
{{\bar k}_1} = \mathop {\arg \max }\limits_{{k_1} \in {\Delta _1}}
\left\{ {\mathop {\max }\limits_{{p_{n{k_1}}} \ge 0}
L_{n{k_1}}^{\lambda\mu} } \right\},
\end{equation}
where $ L_{_{n{k_1}}}^{\lambda \mu } = \frac{{{w_{{k_1}}}}}{N}{\log
_2}\left( {1 + \frac{{{h_{n{k_1}}}}}{{I_{21}^n}}{p_{n{k_1}}}}
\right) - \left( {\lambda  + \frac{{\mu {g_{n{{\bar k}_2}}}}}{N}}
\right){p_{n{k_1}}}, $ with $I_{21}^n = {p_{n{{\bar
k}_2}}}{g_{n{k_1}}} + {\sigma ^2}$ being the interference-plus-noise
power at SC $n$. 
Equ. (\ref{equ.27}) means SC $n$ should be assigned to the user,
denoted by ${{\bar k}_1}$, giving the highest value of
$L_{n{k_1}}^{\lambda\mu}$ with the optimized ${p_{n{k_1}}}$. By
letting ${\partial L_{_{n{k_1}}}^{\lambda \mu }/\partial
{p_{n{k_1}}}}$ be zero and considering non-negative power
allocation, the power allocation in (\ref{equ.27}) should be
optimized according to
\begin{equation}\label{equ.28} {p_{n{k_1}}} = {\left(
{\frac{{{w_{{k_1}}}}}{{N\ln 2}}\frac{1}{{\lambda  + \mu
\frac{{{g_{n{{\bar k}_2}}}}}{N}}} -
\frac{{I_{21}^n}}{{{h_{n{k_1}}}}}} \right)^ + }.
\end{equation}
Thus, (\ref{equ.27}) and (\ref{equ.28}) together provide the optimal
resource allocation rules at all SCs with fixed $\lambda$ and $\mu$.

Then, the ellipsoid method can 
be adopted to iteratively search
over $\lambda\geq 0$ and $\mu\geq 0$ so that the constraints in
(\ref{equ.21}) and (\ref{equ.22}) are simultaneously satisfied. 
This algorithm also bears a linear complexity order i.e.,
$O(ItrK_1N)$, where $Itr$ denotes the number of iterations for
updating $\lambda$ and $\mu$. Similarly as (\ref{equ.20}), the
resource allocation problem for the 2nd cell can be formulated to
maximize $R_2$ subject to the transmit power constraint
$P_2^{\rm{BS}}$ of ${\rm BS}_2$ and the JSP constraint $T_1$ to
protect the active users in the 1st cell. For a given pair of
$T_1\geq 0$ and $T_2\geq 0$,\footnote{The methods for properly
setting $T_1$ and $T_2$ can be found in the journal version of this
paper \cite{Zhang2011TWC}.} the per-cell resource allocation
described above can be iteratively implemented between ${\rm BS}_1$
and ${\rm BS}_2$ until the SC and power allocation in both cells
converges.
\subsection{Individual Subcarrier Protection (ISP)}

In this subsection, we study the decentralized resource allocation
with ISP. Similarly to the previous case of JSP, we merely present
the solution to the optimization problem for the 1st cel. 
With the same objective function as (\ref{equ.20}) and BS transmit
power constraint as (\ref{equ.21}), 
we formulate the current problem via replacing the JSP 
constraint 
in (\ref{equ.22}) 
by the following ISP constraint over each individual SC:
\begin{equation}\label{equ.30}
\sum\nolimits_{{k_1} = 1}^{{K_1}} {{p_{n{k_1}}}{g_{n{{\bar k}_2}}}}
\le T_2^n,{\rm{ }} n \in \Lambda ,
\end{equation}
where $T_2^n$ is the interference power constraint for protecting
the active user at SC $n$ in the 2nd cell.


Again, we apply the Lagrange duality to solve the per-cell resource
allocation problem with ISP. Following a similar procedure as in
JSP, we can derive the following optimal SC and power allocation
rules:
\begin{equation}\label{equ.35}
{{\bar k}_1} = \mathop {\arg \max }\limits_{{k_1} \in {\Delta _1}}
\left\{ {\mathop {\max }\limits_{0\le {p_{n{k_1}}} \le
T_2^n/{g_{n{{\bar k}_2}}}} L_{n{k_1}}^\lambda } \right\},
\end{equation}
\begin{equation}\label{equ.37}
{p_{n{k_1}}} = \min \left\{ {{{\left( {\frac{{{w_{{k_1}}}}}{{N\ln
2}}\frac{1}{\lambda } - \frac{{I_{21}^n}}{{{h_{n{k_1}}}}}} \right)}^
+ },\frac{{T_2^n}}{{{g_{n{{\bar k}_2}}}}}} \right\},
\end{equation}
where $L_{n{k_1}}^\lambda  = \frac{{{w_{{k_1}}}}}{N}{\log _2}\left(
{1 + \frac{{{h_{n{k_1}}}}}{{I_{21}^n}}{p_{n{k_1}}}} \right)  -
\lambda {p_{n{k_1}}}$, with $\lambda$ being the non-negative dual
variable associated with (\ref{equ.30}).

According to (\ref{equ.35}) and (\ref{equ.37}), 
the optimal SC and power allocation can be determined
for all $n$'s
with any given $\lambda\geq 0$. Then, the bisection method
\cite{Boy04} can be used to adjust $\lambda$ so that the BS transmit
power constraint (\ref{equ.21}) is satisfied. 

Nevertheless, it is not computationally efficient to individually
optimize $T_2^n$ ($T_1^n$) for each SC, thus two special schemes are
further identified. One scheme is to set $T_1^n = T_2^n = + \infty,
\forall n \in \Lambda$, which means that each cell is not aware of
its interference to the adjacent cell, named as the \textbf{No
Protection} scheme. The other scheme is to set uniform peak
interference power constraints over all SCs, i.e., $T_1^n = T_1^*,
T_2^n = T_2^*, \forall n \in \Lambda$, named as the \textbf{Peak}
scheme.

\section{Simulation Results}\label{section5}

\begin{figure}[!t]
  \centering
  \includegraphics[width=78mm]{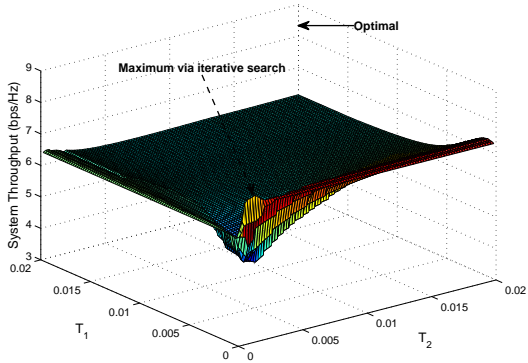} \\
  \vspace{-12pt}
  \caption{System throughput of the decentralized scheme with JSP: One random channel realization.}
  \label{fig.3} \vspace{-12pt}
\end{figure}

\begin{figure}[!t]
  \centering
  \includegraphics[width=78mm]{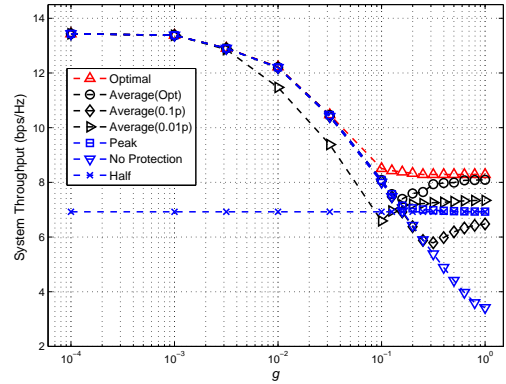} \\
  \vspace{-12pt}
  \caption{System throughput vs. inter-cell interference channel gain.}
  \label{fig.6} \vspace{-8pt}
\end{figure}

\begin{figure}[!t]
  \centering
  \includegraphics[width=78mm]{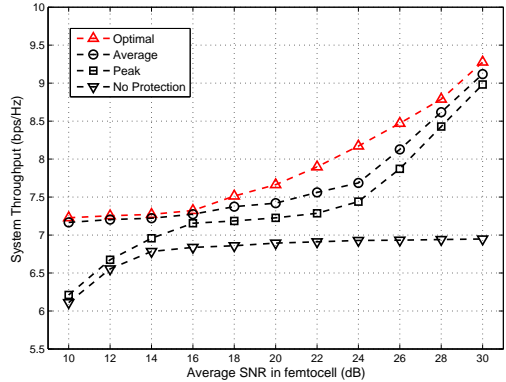} \\
  \vspace{-12pt}
  \caption{System throughput vs. average femtocell SNR.}
  \label{fig.8} \vspace{-8pt}
\end{figure}

In this section, simulation results are presented to evaluate the
performance of the proposed schemes for the two-cell downlink OFDMA
system. It is assumed that $B=100$ MHz and $N=64$. In addition, all
users' rate weights are assumed to be one, and the noise power
spectral density $z_0$ is set to be $-100$ dBm/Hz. Assuming
independent (time-domain) Rayleigh fading with six independent,
equal-energy multipath taps, the frequency-domain channel gains
\{${h_{n{k_1}}}$\}, \{${h_{n{k_2}}}$\}, \{${g_{n{k_1}}}$\}, and
\{${g_{n{k_2}}}$\} are modeled as independent CSCG random variables
distributed as $ \mathcal{CN}\left( {0,a} \right)$, $
\mathcal{CN}\left( {0,b} \right)$, $ \mathcal{CN}\left( {0,c}
\right)$ and $ \mathcal{CN}\left( {0,d} \right)$, respectively. For
convenience, we normalize $a=1$, and adjust $b,c$ and $d$ to
generate different channel models. Figs. \ref{fig.3} and \ref{fig.6}
show the results for two macrocells with $K_1=K_2=8$ and Fig.
\ref{fig.8} for the case with one macrocell and one femtocell with
$K_1=8$ and $K_2=2$ (cf. Fig. 1).

Fig. \ref{fig.3} shows the system throughput, $R_1+R_2$, achieved by
different interference power constraints $T_1$ and $T_2$ using the
proposed decentralized scheme with JSP (i.e., the Average scheme) in
Section IV.A for one particular channel realization. The channel
gains are obtained by setting $b=1$ and $c=d=0.2$, while the
transmit power limits at two BSs are set equally to be $P^{\rm
BS}=1$ watt. 
In this figure, we have marked one local maximum point
obtained by the iterative search method in \cite{Zhang2011TWC}. 
Also, we have marked the system throughput obtained by the
centralized scheme proposed in Section III ({\bf Optimal} scheme).
It is observed that the system throughput achieved by the
decentralized Average scheme is suboptimal as compared to
that by the centralized Optimal scheme. 


Fig. \ref{fig.6} shows the system throughput against the average
inter-cell interference channel gain for various 
schemes. 
The channels are generated via $b=1$ and $c=d=g$, with $g$ being the
average interference channel gain ranging from $10^{-4}$ to 1. 
The proposed decentralized Average scheme 
achieves the system throughput close to that by the centralized
Optimal scheme for all values of $g$, when the searched optimized 
values of $T_1$ and $T_2$ are applied \cite{Zhang2011TWC}. If
instead the preassigned values for $T_1$ and $T_2$ are applied,
throughput degradations are observed to be negligible in the case of
$T_1=T_2=0.1p$ for the low inter-cell interference regime with small
values of $g$, and in the case of $T_1=T_2=0.01p$ for the high
inter-cell interference regime with large values of $g$. 
In addition, the \textbf{Half} scheme (
each cell orthogonally uses half of the overall frequency band) 
and the No Protection scheme are observed to perform poorly for
small and large values of $g$. 
Moreover, 
the Average scheme with JSP performs
superior over the Peak scheme with ISP, 
especially when $g$ becomes large. 


Finally, Fig. \ref{fig.8} shows the system throughput for a
macrocell with a femtocell inside it. 
The channel gains are 
$b=5$, $c=0.1$, and $d=0.5$. The transmit power constraint at the
macrocell's BS is assumed to be 1 watt, while that at the
femtocell's BS is changed from 0.02 to 2 watts.
It is observed that all proposed centralized and decentralized
resource allocation schemes outperform the No Protection scheme in
the achievable system throughput, which eventually becomes saturated
with the increased inter-cell interference. At low femtocell SNR,
there exists a noticeable throughput gap between the Average and
Peak schemes, which is due to the fact that when the femtocell
suffers detrimental interference from the macrocell, the Average
scheme can opportunistically allocate the femtocell transmit power
to a small portion of SCs with best channel conditions. On the other
hand, at high femtocell SNR, both Average and Peak schemes tend to
perform close to the Optimal scheme.



\section{Conclusion}\label{section6}

In this paper, the downlink cooperative interference control in a
two-cell OFDMA system is investigated with centralized and
decentralized implementations for joint power and subcarrier
allocation to maximize the system throughput. It is shown that the
proposed decentralized recourse allocation schemes via the new
approach of inter-cell interference power protection achieve a
performance close to that of the centralized scheme in various
system settings. In addition, the joint subcarrier protection (JSP)
with average interference power constraint is shown to achieve a
larger system throughput than the more stringent individual
subcarrier protection (ISP) counterpart with peak interference power
constraint. 

\begin{thebibliography}{1}

\bibitem{Gesb10}
D. Gesbert, S. Hanly, H. Huang, S. Shamai Shitz, O. Simeone, and W.
Yu, ``Multi-cell MIMO cooperative networks: A new look at
interference,'' \emph{IEEE J. Select. Areas Communications}, vol.
28, no. 9, pp. 1380-1408, Dec. 2010.

\bibitem{Vent09}
L. Venturino, N. Prasad, and X. Wang, ``Coordinated scheduling and
power allocation in downlink multicell OFDMA networks,'' \emph{IEEE
Trans. Vehicular Technology}, vol. 58, no. 6, pp. 2835-2848, July
2009.

\bibitem{Yu10}
W. Yu, T. Kwon, and C. Shin, ``Joint scheduling and dynamic power
spectrum optimization for wireless multicell networks,'' in
\emph{Proc. 44th Conf. Info. Science Sys.} (\emph{CISS}), Princeton,
NJ, March 2010, pp. 1-6.

\bibitem{Zhang10new}
R. Zhang and S. Cui, ``Cooperative interference management with MISO
beamforming,'' \emph{IEEE Trans. Signal Processing}, vol. 58, no.
10, pp. 5450-5458, Oct. 2010.

\bibitem{Boy04}
S. Boyd and L. Vandenberghe, Convex Optimization. Cambridge, U.K.:
Cambridge Univ. Press, 2004.

\bibitem{Yu05}
W. Yu and R. Liu, ``Dual methods for nonconvex spectrum optimization
of multicarrier systems,'' \emph{IEEE Trans. Communications}, vol.
54, no. 7, pp. 1310-1322, July 2006.

\bibitem{Zhang2011TWC}
B. Da and R. Zhang, ``Cooperative interference control for spectrum
sharing in cellular OFDMA systems,'' submitted for publication.






























\end{thebibliography}
\end{document}